\documentclass[aps,prl,reprint,superscriptaddress,amsmath,footinbib]{revtex4-1}
\usepackage{graphicx}
\usepackage{bbold}
\usepackage{hyperref}
\usepackage{import}
\usepackage{color}
\usepackage{soul}
\usepackage{mathrsfs}
\usepackage{amsbsy}

\newcommand{\ket}[1]{\ensuremath{\,|#1\rangle}}

\begin{document}

\title{Ion-based nondestructive sensor for cavity photon numbers}
\author{Moonjoo Lee}
\affiliation{Institut f\"ur Experimentalphysik, Universit\"at Innsbruck, Technikerstra\ss e 25, 6020 Innsbruck, Austria}
\author{Konstantin Friebe}
\affiliation{Institut f\"ur Experimentalphysik, Universit\"at Innsbruck, Technikerstra\ss e 25, 6020 Innsbruck, Austria}
\author{Dario A. Fioretto}
\affiliation{Institut f\"ur Experimentalphysik, Universit\"at Innsbruck, Technikerstra\ss e 25, 6020 Innsbruck, Austria}
\author{Klemens Sch\"{u}ppert}
\affiliation{Institut f\"ur Experimentalphysik, Universit\"at Innsbruck, Technikerstra\ss e 25, 6020 Innsbruck, Austria}
\author{Florian R. Ong}
\affiliation{Institut f\"ur Experimentalphysik, Universit\"at Innsbruck, Technikerstra\ss e 25, 6020 Innsbruck, Austria}
\author{David Plankensteiner}
\affiliation{Institut f\"ur Theoretische Physik, Universit\"at Innsbruck, Technikerstra\ss e 21, 6020 Innsbruck, Austria}
\author{Valentin Torggler}
\affiliation{Institut f\"ur Theoretische Physik, Universit\"at Innsbruck, Technikerstra\ss e 21, 6020 Innsbruck, Austria}
\author{Helmut Ritsch}
\affiliation{Institut f\"ur Theoretische Physik, Universit\"at Innsbruck, Technikerstra\ss e 21, 6020 Innsbruck, Austria}
\author{Rainer Blatt}
\affiliation{Institut f\"ur Experimentalphysik, Universit\"at Innsbruck, Technikerstra\ss e 25, 6020 Innsbruck, Austria}
\affiliation{Institut f\"ur Quantenoptik und Quanteninformation, \"Osterreichische Akademie der Wissenschaften, Technikerstra\ss e 21a, 6020 Innsbruck, Austria}
\author{Tracy E. Northup}
\email[Corresponding author: ]{tracy.northup@uibk.ac.at}
\affiliation{Institut f\"ur Experimentalphysik, Universit\"at Innsbruck, Technikerstra\ss e 25, 6020 Innsbruck, Austria}

\date{\today}


%%%%% 0. Abstract
\begin{abstract}
We dispersively couple a single trapped ion to an optical cavity to extract information about the cavity photon-number distribution in a nondestructive way. 
The photon-number-dependent AC-Stark shift experienced by the ion is measured via Ramsey spectroscopy.
We use these measurements first to obtain the ion-cavity interaction strength. 
Next, we reconstruct the cavity photon-number distribution for coherent states and for a state with mixed thermal-coherent statistics, finding overlaps above 99\% with the calibrated states.
\end{abstract}

\maketitle
% Introduction

Cavity quantum electrodynamics (cavity QED) provides a conceptually simple and powerful platform for probing the quantized interaction between light and matter~\cite{Haroche06}. 
Early experiments opened a window into the dynamics of coherent atom--photon interactions, first through observations of collective Rabi oscillations and vacuum Rabi splittings~\cite{Kaluzny83, Raizen89, Bernardot92, Brecha95} and later at the single-atom level~\cite{Thompson92,Brune96,Childs96,Hood98,Boca04,Maunz05}.
More recently, building on measurements of the cavity field via the atomic phase \cite{Brune94,Bertet02}, cavity photon statistics have been analyzed in experiments with Rydberg atoms or superconducting qubits in microwave resonators~\cite{Schuster05, Schuster07, Guerlin07, Johnson10}, culminating in the generation and stabilization of nonclassical cavity field states~\cite{Deleglise08, Hofheinz09, Sayrin11, Vlastakis13, Heeres15, Holland15, Wang16}. 
These experiments operate in a dispersive regime, in which information about the cavity field can be extracted via the qubits with minimal disturbance to the field~\cite{Haroche06}.
	
In parallel, it was pointed out that the Jaynes-Cummings Hamiltonian that describes cavity QED also describes the interaction of light and ions in a harmonic trapping potential~\cite{Blockley92}.
This interaction underpins the generation of nonclassical states of motion~\cite{Cirac93,Meekhof96,Monroe96,Kienzler15} and the implementation of gates between trapped ions~\cite{Haeffner08}.
In analogy to the phase shifts experienced by qubits due to the cavity field, ions experience quantized AC-Stark shifts due to their coupling to the harmonic trap potential~\cite{Schmidt-Kaler04}. These shifts have been characterized using techniques similar to those introduced in Ref.~\cite{Brune94}.
Here, we have transferred the principle of dispersive measurement to an ion qubit coupled to a cavity.  
In contrast to experiments with flying Rydberg atoms, the ion is strongly confined; in contrast to both Rydberg and superconducting-qubit experiments, our cavity operates in the optical regime.

%%%%% 3. Basic idea
We employ a single trapped $^{40}$Ca$^{+}$ ion as a quantum sensor~\cite{Degen17} to extract information about cavity photons without destroying them.
Via Ramsey spectroscopy of the ion, we measure the phase shift and dephasing of the ion's state, both of which result from the interaction of the ion with the cavity field. 
The phase shift is induced by the mean number of cavity photons due to the AC-Stark effect, and the dephasing is caused by uncertainties in the cavity photon number.
Reconstructing the cavity photon-number distribution from these measurements allows us to determine the mean and the width of the distribution and thus to distinguish between states with coherent photon statistics and mixed thermal-coherent statistics.

%%%%% 4. Setup and energy levels
The ion is modelled as a three-level system in which two states, $\ket{S}\equiv\ket{4^2\text{S}_{1/2},m_J=+1/2}$ and $\ket{D}\equiv\ket{3^2\text{D}_{5/2},m_J=+1/2}$, comprise a qubit (Fig.~1). 
The cavity is dispersively coupled to the transition between $|D\rangle$ and the third state, $\ket{P}\equiv\ket{4^2\text{P}_{3/2},m_J=+1/2}$, with a detuning $\Delta = 2\pi \times 125$~MHz.
The quantization axis is defined by a magnetic field of $4.06$~G in the plane perpendicular to the cavity axis.
The relevant ion-cavity parameters are given by $(g, \kappa, \gamma) = 2\pi \times$(0.968, 0.068, 11.5) MHz, where $g$ is the ion-cavity coupling strength calculated from the cavity properties and the atomic transition, $\kappa$ is the cavity field decay rate, and $\gamma$ is the atomic decay rate of state $\ket{P}$. 
Here, we assume that the ion is positioned at the waist and in an antinode of a TEM$_{00}$ mode of the cavity~\cite{Russo09,Stute12a}.
The expected frequency shift of the cavity resonance induced by the dispersively coupled ion is $g^2/\Delta = 2\pi\times7.50$~kHz, which is much smaller than $\kappa$, such that we operate in the weak-pull regime~\cite{Gambetta06, Murch13}.
In this regime, the drive laser can be considered to be resonant with the cavity, irrespective of the state of the qubit.

%%%%% 5. Sequence
In order to probe the cavity field, the ion is first Doppler-cooled and optically pumped to $|S\rangle$. 
As the first part of a Ramsey measurement, the qubit is then initialized in a superposition of $|S\rangle$ and $|D\rangle$ by a $\pi/2$-pulse of the Ramsey spectroscopy laser at $729$~nm. 
Next, we drive the cavity with a weak laser field with wavelength $\lambda_\text{L}=854$~nm for $T=50~\mu$s.
Note that the interaction time $T$ is much larger than the cavity photon lifetime of $\tau_\text{C}=1/(2\kappa) = 1.2~\mu$s, such that for a mean intracavity photon number of $\langle n\rangle$, $\langle n\rangle T/\tau_\text{C}$ photons on average successively interact with the ion.
Note also that $T$ is much shorter than the coherence time of $950~\mu$s on the $|S\rangle$--$|D\rangle$ transition~\cite{Schindler13}.
The independently calibrated mean photon number $\langle n \rangle$ of the cavity field is set to a value between $0$ and $1.6(3)$, and the drive laser frequency $\omega_{\rm{L}}=2\pi c/\lambda_\text{L}$ is resonant with the cavity frequency $\omega_{\rm{C}} + \langle n \rangle g^2/\Delta$, where $\omega_{\rm{C}}$ is the cavity resonance frequency when no ion is coupled to the cavity.
Finally, a second $\pi/2$-pulse with variable phase $\phi$ completes the Ramsey measurement, after which the qubit state is detected using laser fields at $397$~nm and $866$~nm~\cite{Schindler13}.
The measurement is repeated $250$ times for each phase to obtain the ion population in $|D\rangle$.

%%%%% Fig. 1
\begin{figure} [!t]
	\includegraphics[width=3.5in]{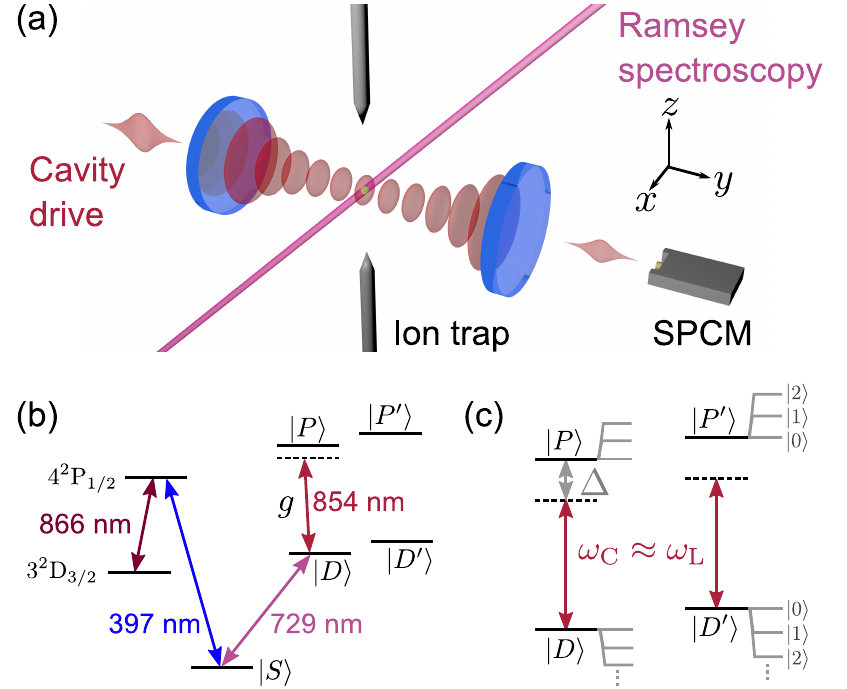} 
	\caption{(a) Experimental set-up. 
		A single ion is coupled to the cavity, which is driven by a weak laser field (cavity drive).
		The cavity drive laser (along $\hat{y}$) is polarized parallel to the quantization axis, in the direction $\hat{x}+\hat{z}$. 
		The Ramsey spectroscopy laser propagates along $-(\hat{y}+\hat{z})$.
		Cavity output photons are detected by a single-photon-counting module (SPCM).
		RF trap electrodes are omitted for clarity.
		(b) Energy level diagram of $^{40}$Ca$^+$ with the relevant levels $\ket{S}$, $\ket{D}$, $\ket{P}$, $\ket{D'}\equiv\ket{3^2\text{D}_{5/2},m_J=+3/2}$	and $\ket{P'}\equiv\ket{4^2\text{P}_{3/2},m_J=+3/2}$ of the ion.  
		The $4^2\text{P}_{1/2}$ and $3^2\text{D}_{3/2}$ manifolds are used for ion cooling and detection.
		(c) Levels $\ket{D}$, $\ket{P}$, $\ket{D'}$, and $\ket{P'}$ experience photon-number-dependent AC-Stark shifts due to the cavity field, indicated in grey. 
		The frequencies of the bare cavity and the drive laser are $\omega_\text{C}$ and $\omega_\text{L}$, respectively, and $\Delta$ is the difference between $\omega_\text{C}$ and the transition frequency from $|D \rangle$ to $|P \rangle$.
		}
	\label{fig:Setup}
\end{figure}

\begin{figure*} [!t]
\includegraphics[width=7in]{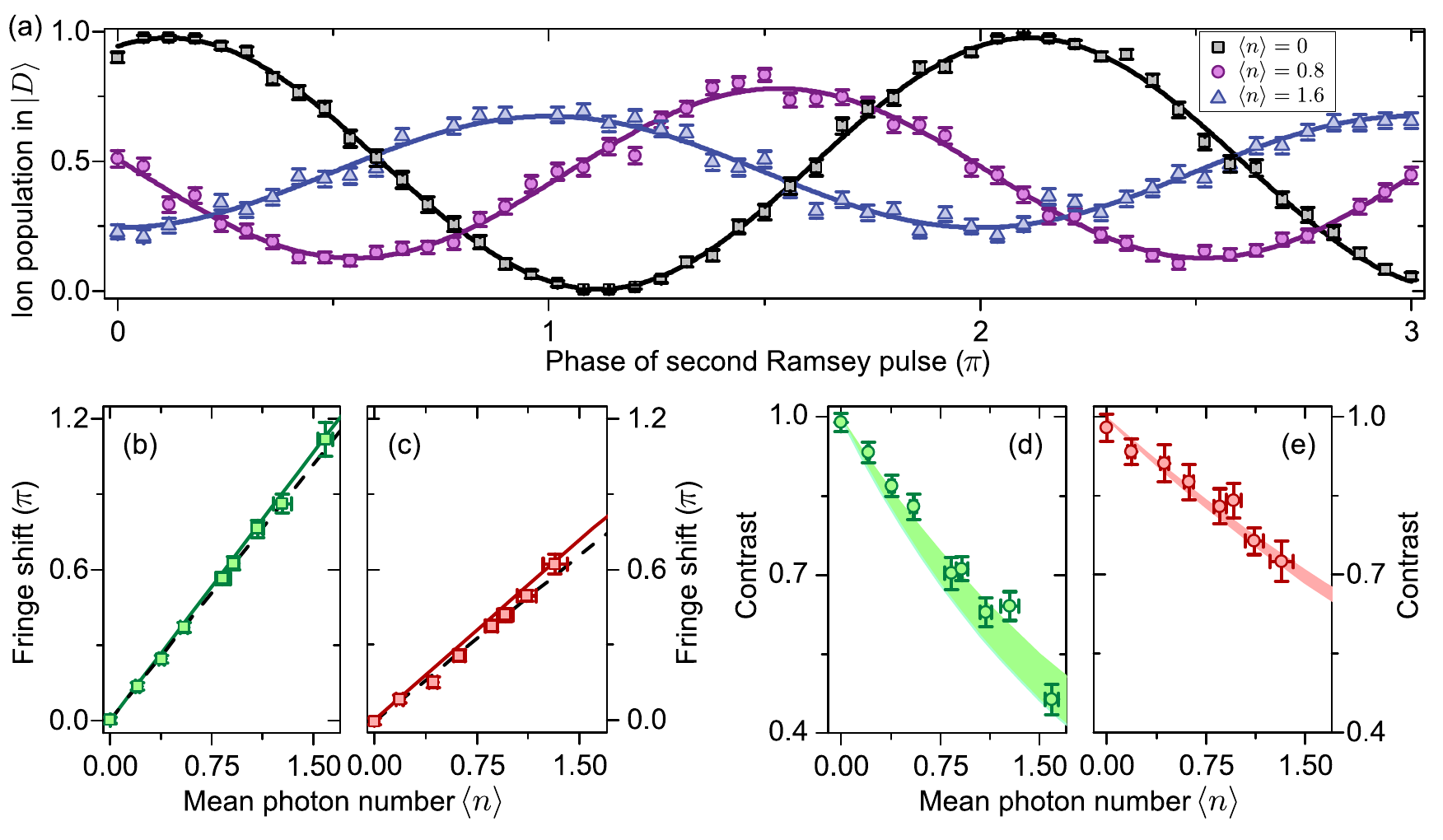} 
	\caption{
		(a)	Ramsey fringes for mean photon numbers $\langle n \rangle =0$ (black squares), $0.8(2)$ (purple circles), and $1.6(3)$ (blue triangles).
			The solid lines are sinusoidal fits~\cite{suppInf} and error bars denote quantum projection noise.
		(b)	The phase shift of the Ramsey fringes as a function of $\langle n \rangle$ for the transition $\ket{D}-\ket{P}$.
			Squares are experimental data, while the solid line shows the theoretical model using the calculated coupling strength $g$. The dashed line is a linear fit to the data, from which $g_\text{exp}$ is extracted (see main text).
		(c) Ramsey fringe phase shift as a function of $\langle n\rangle$ for the transition $\ket{D'}-\ket{P'}$ with $g'=0.82\, g$.
		(d)	Contrast of the Ramsey fringes as a function of $\langle n \rangle$ for the transition $\ket{D}-\ket{P}$.
			The shaded area shows the contrast expected from the theoretical model with $g_\text{exp}$ as input, including its uncertainty.
		(e) Contrast vs. $\langle n\rangle$ for the transition $\ket{D'}-\ket{P'}$.
			For (b)-(e), the plotted uncertainties in $\langle n\rangle$ are statistical uncertainties from the calibration of the photon number. 
			Systematic uncertainties in $\langle n\rangle$ are $20\%$. 
			Error bars of fringe shift and contrast are uncertainties of the fits to the Ramsey fringes.
	}
	\label{fig:Ramsey}	
\end{figure*}

%%%%% 6. Result1: phase shift of Ramsey fringes
The mean population in $|D\rangle$ as a function of the phase $\phi$ is plotted in Fig.~\ref{fig:Ramsey}(a) for three values of $\langle n \rangle$.
As $\langle n \rangle$ is increased, two features emerge: the Ramsey fringe is shifted, and its contrast is reduced. 
The phase shift is directly proportional to $\langle n \rangle$, as shown in Fig.~\ref{fig:Ramsey}(b), with proportionality factor $T g^2/\Delta$.
For $\langle n \rangle = 0.8(2)$ and $1.6(3)$, the phase of the qubit is shifted by $0.57(3)\pi$ and $1.12(7)\pi$, respectively.
A single photon only interacts with the ion during its time in the cavity, which has a mean value $\tau_\text{C}$, corresponding to a phase shift of the ion by $\tau_\text{C}\,g^2/\Delta=0.018\pi$. 
The accumulated effect of all successive photons injected into the cavity accounts for the total phase shift of the qubit.

%%%%% 7. measurement of coupling strength
The measured phase shift as a function of $\langle n\rangle$ can be used to determine the ion-cavity coupling strength.
This method is independent of the single-atom cooperativity and thus is valid also for systems in intermediate and even weak coupling regimes.
In such regimes, observing the vacuum Rabi splitting is not possible, making it difficult to measure the coupling strength directly. 
As we have independently determined all ion-cavity parameters and calibrated the photo-detection efficiency, we fit a theoretical model to the data with the coupling strength as the only free parameter.
In this way, we extract the experimental value of $g_\text{exp}=2\pi\times 0.96(4)$~MHz from the data displayed in Fig.~\ref{fig:Ramsey}(b), in agreement with the theoretical value of $g=2\pi\times0.968~\text{MHz}$.
We performed the same set of measurements on another $^{40}$Ca$^+$ transition, using the states $|S \rangle$, $\ket{D'}\equiv\ket{3^2$D$_{5/2}, m_{J}=+3/2}$, and $\ket{P'}\equiv\ket{4^{2}$P$_{3/2}, m_{J}=+3/2}$ (Fig.~\ref{fig:Ramsey}(c));
the coherence time for the transition $\ket{S}-\ket{D'}$ is $510~\mu\text{s}$.
For the transition $\ket{D'}-\ket{P'}$, we expect $g'=2\pi\times0.790~\text{MHz}$ and extract $g_\text{exp}'=2\pi\times0.77(4)$ MHz. 
From the two independent measurements on two transitions, we thus see that this new method determines the atom-cavity coupling strength in agreement with theory.

%%%%% 8. Result2: contrast reduction of Ramsey fringes
In Fig.~\ref{fig:Ramsey}(d), the fringe contrast, defined as the fringe amplitude divided by the fringe offset, is plotted as a function of $\langle n \rangle$ for the transition $\ket{D}-\ket{P}$ and in Fig.~\ref{fig:Ramsey}(e) for the transition $\ket{D'}-\ket{P'}$.
For $\ket{D}-\ket{P}$, the contrast decreases from $0.99(2)$ to $0.46(3)$ as $\langle n \rangle$ increases from $0$ to $1.6$.
This reduction reflects the fact that the intracavity photon number is inherently probabilistic, and in this case, for a coherent drive, follows a Poissonian distribution.
The corresponding photon-number fluctuations in the cavity field lead to fluctuations of the qubit transition frequency through the photon-number-dependent AC-Stark shift. 
Since in our case $T\gg \tau_\text{C}$, the ion qubit additionally dephases during the interaction.
This dephasing can also be interpreted as the consequence of a weak measurement of the qubit state~\cite{Schuster05}: 
Intracavity photons interact dispersively with the qubit before leaking to the environment. 
The phase of the output photons thus carries information about the qubit state that could be accessed, e.g., with homodyne or heterodyne detection. 
All such quantum measurements imply some amount of backaction~\cite{Gambetta06}, which in our case takes the form of qubit decoherence. 
Note that in the absence of a cavity, photons would also induce an AC-Stark shift of the ion's states, but 
due to the weakness of the free-space interaction, the effect would be too small to be measured at the single-photon level.

%%%%% Fig. 2
\begin{figure}[!t]
	\includegraphics[width=3.5in]{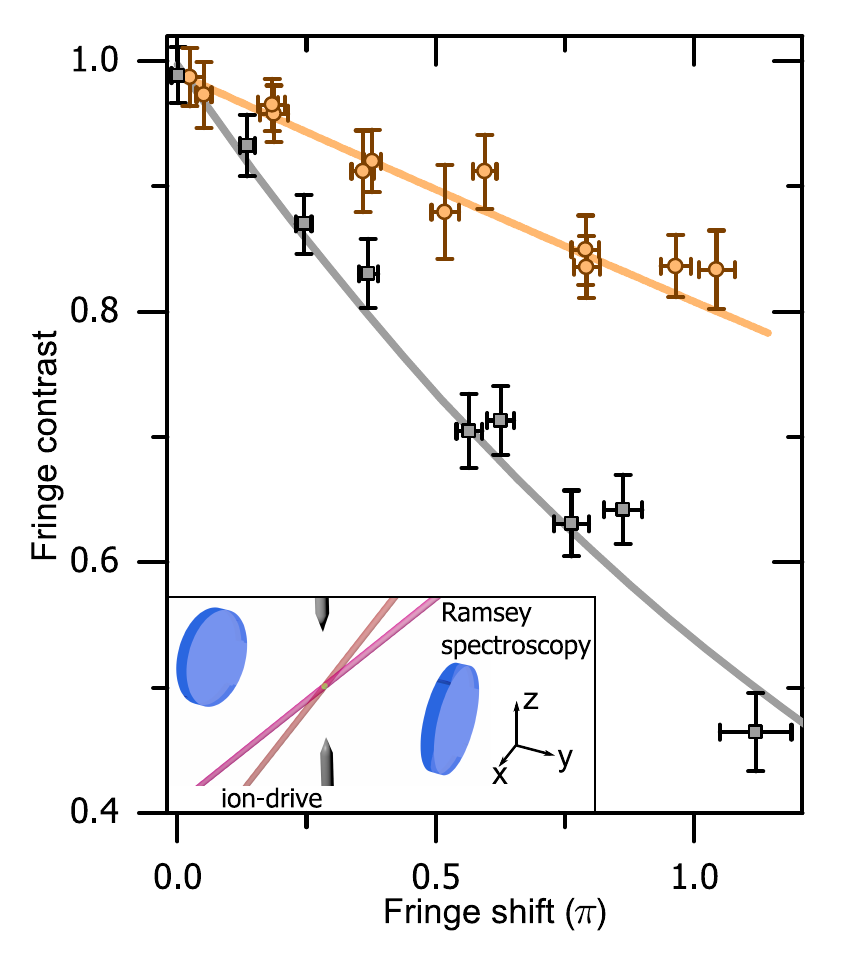} 
	\caption{
		Ramsey fringe contrast as a function of phase shift for ion-drive (orange circles) and cavity-drive (black squares; same data as in Fig.~\ref{fig:Ramsey}(b) and (d)) measurements on the $|D\rangle$--$|P\rangle$ transition. 
		The lines are theory curves, using $g_\text{exp}$ for the cavity-drive data.
		The inset shows the ion-drive beam, which propagates along $\hat{x}-\hat{z}$ and is polarized along $\hat{x}+\hat{z}$, along with the Ramsey spectroscopy beam. 
		The ion is decoupled from the cavity for the ion-drive measurement.
	}
	\label{fig:ion_drive}
\end{figure}

%%%%% 9. Ion drive measurement
Spontaneous emission contributes to decoherence for both the cavity-drive measurement of Fig.~\ref{fig:Ramsey} and free-space measurements. 
We quantify this effect in a reference measurement using an ``ion-drive'' configuration:
The cavity is translated by a few mm along $\hat{x}$ in order to decouple it from the ion.
The ion is driven with a laser beam with frequency $\omega_\text{L}=\omega_\text{C}$. 
We perform Ramsey measurements with the cavity interaction replaced by the interaction of the ion with this ion-drive laser. 
The Ramsey fringe contrast is reduced due to off-resonant excitation of the population from $|D\rangle$ to $|P\rangle$, followed by spontaneous emission.
Fig.~\ref{fig:ion_drive} compares the Ramsey fringe contrast as a function of the phase shift between the ion-drive and cavity-drive measurements. 
A given phase shift corresponds to the same Rabi frequency in both measurements.
The contrast of the cavity-drive data is smaller than that of the ion-drive data 
because in the former case, both spontaneous emission and decoherence induced by the cavity photons play a role.
We therefore conclude from this reference measurement, that a significant contribution to decoherence of the ion qubit in the cavity-drive configuration stems from interaction with the cavity field via the backaction of the cavity photons on the ion and not from spontaneous emission.

%%%%% Fig. 3

%%%%% Fig. 4
\begin{figure}[!t]
	\includegraphics[width=3.2in]{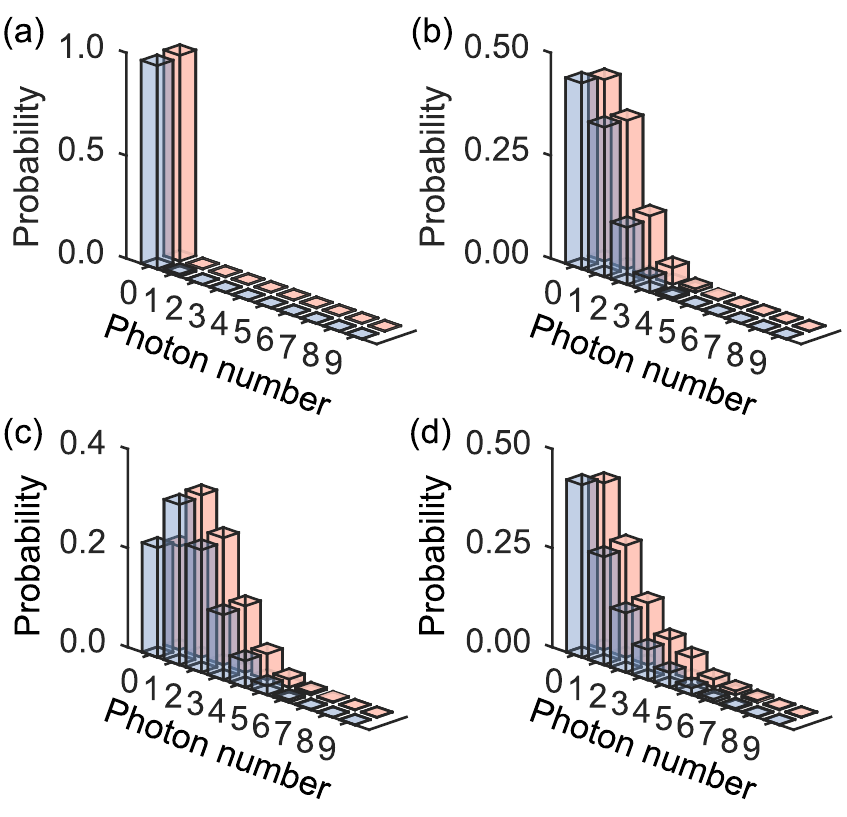} 
	\caption{
		Photon number distributions reconstructed from the measured Ramsey fringes for intracavity mean photon numbers of (a) $0$, (b) $0.8(2)$, and (c) $1.6(3)$ (blue bars), and the expected distributions (pink bars).
		The reconstructed distributions yield mean photon numbers of $0.01^{+0.05}_{-0.02}$, 		$0.84(8)$, and $1.49^{+0.05}_{-0.06}$. 
		(d) Reconstructed distribution when the cavity is driven with light of mixed coherent-thermal statistics with mean photon number $\langle n\rangle=1.05^{+0.07}_{-0.11}$, yielding a reconstructed mean photon number of $\langle n\rangle=1.12^{+0.14}_{-0.15}$.
		The squared statistical overlap between the reconstructed distributions and the expected distributions is higher than $0.99$ for (a)-(d).
	}
	\label{fig:reconstruction}
\end{figure}

%%%%% 10. Reconstruction
Next, we reconstruct the cavity photon number distribution with a maximum likelihood algorithm~\cite{suppInf}. 
This algorithm finds the photon number distribution that is most likely to have interacted with the ion.
It is based on a model, in which the coherent cavity drive with mean photon number $n_\text{coh}$ is described by an amplitude $\eta=\kappa\sqrt{n_\text{coh}}$, and additional number fluctuations are described by a thermal bath with mean photon number $n_\text{th}$ corresponding to an incoherent contribution to the driving~\cite{Gardiner04}.
The photon number distribution of the intracavity field is then determined by the two parameters $\eta$ and $ n_\text{th}$. 
The result of the reconstruction is shown in Fig.~\ref{fig:reconstruction}.
For the three Ramsey fringes measured on the $|D\rangle-|P\rangle$ transition, displayed in Fig.~\ref{fig:Ramsey}(a), the reconstruction yields a squared statistical overlap (SSO) $\left(\sum_{n}\sqrt{p_\text{rec}(n)p_\text{cal}(n)}\right)^2$ between the reconstructed distribution $p_\text{rec}(n)$ and the independently calibrated input state distribution $p_\text{cal}(n)$ above 99\% (Figs.~\ref{fig:reconstruction}(a)-(c)).
The reconstructed state shown in Fig.~\ref{fig:reconstruction}(a) corresponds to the vacuum state, and the states in Fig.~\ref{fig:reconstruction}(b) and (c) are coherent states, with Mandel $Q$ parameters $Q=\left(\left<n^2\right>-\left<n\right>^2\right)/\left<n\right>-1$ of $0.00^{+0.02}_{-0.01}$, $-0.03(7)$, and $0.04(5)$, respectively~\cite{Glauber-OCPS07}.
The uncertainty of the reconstructed distribution is dominated by quantum projection noise in the Ramsey measurement~\cite{suppInf}.

This reconstruction method is also applied to a fourth state which is generated by applying amplitude noise to the cavity drive laser via an acousto-optic modulator.
The noise has a bandwidth of $10~\text{MHz}\gg 2\kappa$ and can therefore be considered as white noise.
The reconstructed state, shown in Fig.~\ref{fig:reconstruction}(d), can be described by mixed coherent and thermal statistics:
From the calibration of the added noise~\cite{suppInf}, a value of $Q=0.64(6)$ is expected, while the reconstruction yields $Q=0.70^{+0.07}_{-0.10}$.
The result thus shows super-Poissonian intracavity photon statistics caused by the added thermal noise and is clearly distinct from the statistics of a coherent state.
Note that our sensing technique is nondestructive because the dispersive interaction with the ion does not annihilate the measured intracavity photons.

%%%%% 11. Arbitrary states and off-diagonal components
An extension of this work would be to reconstruct the full density matrix of arbitrary states of the cavity field.
For this purpose, we require a displacement operation of the cavity field, as has been demonstrated in microwave cavities~\cite{Deleglise08}.
With the target field to be measured populating the cavity, a second field as a local oscillator would be sent to the cavity.
The total field interacting with the ion would be the sum of the known (local oscillator) and unknown  (target) fields, and by varying the known field and measuring the state of the ion, one would be able to extract the full target field density matrix.

%%%%% 11. Outlook and applications
We have focused here on measuring the ion's state to extract information about the cavity field. 
However, the scenario can be reversed:
quantum nondemolition measurements of the ion's state become possible in our setup via heterodyne measurement of the cavity output field, allowing single quantum trajectories of the ion's electronic state to be monitored and the qubit state to be stabilized, as demonstrated with superconducting qubits~\cite{Vijay12,Murch13}. 
Furthermore, the strong-pull regime ($g^2/\Delta > \kappa$) would be accessible with a higher finesse cavity~\cite{Gambetta06,Murch13,suppInf}.
In this regime, the qubit spectrum splits into several lines, each corresponding to a different photon-number component~\cite{Schuster07,Mekhov07}, providing a route to engineer nonclassical cavity-field states in the optical domain. 
Other possible extensions include increasing the sensitivity of the measurement by using several ions via their collective coupling to the cavity~\cite{Begley16} or via their entanglement~\cite{Leibfried05}. 

%%%%% 12. Conclusion
In summary, we have implemented an ion-based analyzer for the statistics of optical photons that does not destroy the photons.
Information about the intracavity photon number is imprinted onto the state of an ion qubit via a dispersive interaction. 
Ramsey spectroscopy and the maximum likelihood method are used to reconstruct the intracavity photon statistics, yielding results in excellent agreement with the expected distributions.
Our work represents the first such nondestructive probing of cavity photon distributions in the optical domain, providing tools for the generation of nonclassical optical states.

%%%%% 13. Acknowledgements, funding 
This work has been financially supported by the Austrian Science Fund (FWF) through Projects F4019, V252, M1964, W1259-N27, and F4013-N23; by the Army Research Laboratory's Center for Distributed Quantum Information via the project SciNet: Scalable Ion-Trap Quantum Network, Cooperative Agreement No. W911NF15-2-0060; and by the European Union's Horizon 2020 research program through the Marie Sk{\l}odowska-Curie Actions, Grant No. 656195.

M.L. and K.F. contributed equally to this work. 

%\bibliography{bibliography}

%

%%%%%%%%%% Merge with supplemental materials %%%%%%%%%%
%\pagebreak
\clearpage
\twocolumngrid
%\widetext
\begin{center}
\textbf{\large Supplemental Material: Ion-based nondestructive sensor for cavity photon numbers}
\end{center}
%%%%%%%%%% Merge with supplemental materials %%%%%%%%%%
%%%%%%%%%% Prefix a "S" to all equations, figures, tables and reset the counter %%%%%%%%%%
\setcounter{equation}{0}
\setcounter{figure}{0}
\setcounter{table}{0}
\setcounter{page}{1}
\makeatletter
\renewcommand{\theequation}{S\arabic{equation}}
\renewcommand{\thefigure}{S\arabic{figure}}
\renewcommand{\bibnumfmt}[1]{[S#1]}
\renewcommand{\citenumfont}[1]{S#1}
%%%%%%%%%% Prefix a "S" to all equations, figures, tables and reset the counter %%%%%%%%%%

%\section{Appendix}
%\appendix

%%%% Describing the system
\section{Modelling the system}
% Atomic levels
\subsection{Atomic levels}
\begin{figure*}[!t]
	\includegraphics[width=6.0in]{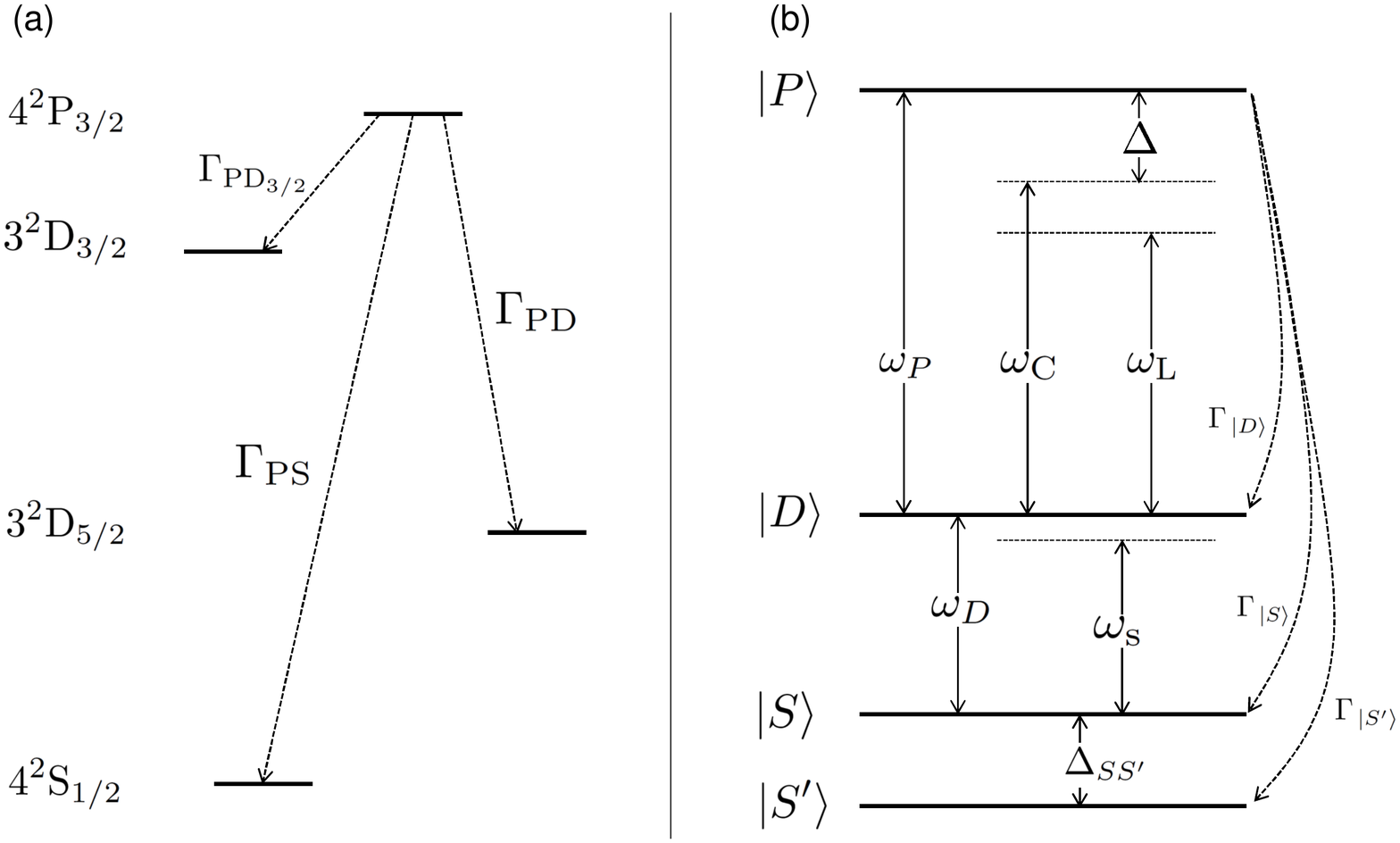} 
	\caption{Definition of the involved levels and transitions. (a) Levels and decay channels considered in the model. 
		Note that decay to $4^2$S$_{1/2}$ from $3^2$D$_{5/2}$ is not included, since the lifetime of the latter level of $1$~s is much longer than the duration of one experimental cycle. 
		(b) Definition of the levels, energies, detunings and decay constants in the model. Note that the decay from $\ket{P}$ to $\ket{S'}$ combines decay channels ending in $4^2$S$_{1/2}$, $3^2$D$_{5/2}$, and $3^2$D$_{3/2}$.}
	\label{fig:levels} 
\end{figure*}

In order to calculate the theory lines in Fig.~2(b)-(e) in the main text, we consider the following atomic basis states: 
	$|S\rangle=|4^2$S$_{1/2}, m_{J}=+1/2 \rangle$, $|D\rangle=|3^2$D$_{5/2}, m_{J}=+1/2\rangle$, $|P\rangle=|4^{2}$P$_{3/2}, m_{J}=+1/2\rangle$, 
	and $|S'\rangle$ (see Fig.~\ref{fig:levels}), where $|S'\rangle$ is a dark state, which collects spontaneous emission from $|P\rangle$ to the second ground state $|4^2$S$_{1/2}, m_{J}=-1/2 \rangle$, as well as to the states $|3^2$D$_{5/2}, m_{J}=-1/2 \rangle$, $|3^2$D$_{5/2}, m_{J}=+3/2 \rangle$, and the states in the $3^2\text{D}_{3/2}$ manifold.		
	$\ket{S'}$ does not participate in the ion-cavity interaction and is not coupled to the Ramsey spectroscopy laser. 
The manifolds involved in the process are displayed in Fig.~\ref{fig:levels}(a). 
The total decay rate of state $\ket{P}$ is $\Gamma_P =\Gamma_\text{PS} + \Gamma_\text{PD} + \Gamma_{\text{PD}_{3/2}}= 2\pi\times 23~\text{~MHz}$, with values $\Gamma_\text{PS}=2\pi\times 21.4$~MHz for decay from $4^2\text{P}_{3/2}$ to $4^2\text{S}_{1/2}$, $\Gamma_\text{PD}=2\pi\times 1.34$~MHz for decay from $4^2\text{P}_{3/2}$ to $3^2\text{D}_{3/2}$, and $\Gamma_{\text{PD}_{3/2}}=2\pi\times0.152$~MHz for decay from $4^2\text{P}_{3/2}$ to $3^2\text{D}_{3/2}$.
Taking into account the Clebsch-Gordan coefficients, the decay rates are $\Gamma_{\ket{S}}=2/3\,\Gamma_\text{PS} = 2\pi\times 14.3~\text{MHz}$ from $\ket{P}$ to $\ket{S}$,  $\Gamma_{\ket{S'}}=1/3\,\Gamma_{PS} + 3/5\,\Gamma_{PD} + \Gamma_{PD_{3/2}} = 2\pi\times8.1~\text{MHz}$ from $\ket{P}$ to $\ket{S'}$, and $\Gamma_{\ket{D}} = 2/5\,\Gamma_{PD} = 2\pi\times 0.54~\text{MHz}$ from $\ket{P}$ to $\ket{D}$.

% Fig. S1

% Hamiltonian
\subsection{Hamiltonian}
The Hamiltonian of the system is given by
	\begin{align}
		H_{S}/\hbar= & \omega_{D}\sigma_{D} + \left(\omega_{D} + \omega_P\right)\sigma_{P}   \\ 
		& + \omega_\text{C}a^{\dagger}a + g\left(\sigma_{PD}a + \text{h.c.}\right)\nonumber \\
		&  + \left(\eta a^{\dagger} e^{-i\omega_\text{L}t} + \text{h.c.} \right) \nonumber  +\left(\Omega \sigma_{SD} e^{-i\omega_\text{R}t} + \text{h.c.}\right). \nonumber
	\end{align}
Here, $\omega_D$ corresponds to the energy of the level $\ket{D}$, $\omega_P$ to that of $\ket{P}$, $\sigma_{D(P)}$ is the projection operator onto the state $\ket{D}$ ($\ket{P}$), $\omega_\text{C}$ is the cavity frequency, $a$ is the annihilation operator of the cavity mode, $g$ is the ion-cavity coupling strength, $\sigma_{PD}=\sigma_{DP}^\dagger$ is the transition operator between states $\ket{P}$ and $\ket{D}$, $\eta$ is the amplitude of the drive laser in the cavity drive term, $\omega_\text{L}$ is the frequency of the cavity drive laser, $\Omega$ is the Rabi frequency of the laser on the $\ket{S}$--$\ket{D}$ qubit transition, $\sigma_{SD}=\sigma_{DS}^\dagger$ is the transition operator between states $\ket{S}$ and $\ket{D}$, and $\omega_\text{R}$ is the frequency of the Ramsey spectroscopy laser. 
The energy of the ground state $\ket{S}$ is chosen as the energy reference.
Fig.~\ref{fig:levels}(b) shows the relevant states, frequencies and decay channels.
This Hamiltonian is transformed into a rotating frame via 
	\begin{align*}
		H_{I}=i\hbar\dot{U}U^{\dagger}+UH_{S}U^{\dagger}, 
	\end{align*}
with a unitary operator
	\begin{align}
		U = \exp\left[ i \left(\omega_\text{R} \sigma_{D} + (\omega_s+\omega_P)\sigma_{P} + \omega_P a^\dagger a  \right)t\right]. \nonumber
	\end{align}
We thus obtain the Hamiltonian in the interaction picture as
	\begin{align}
		H_I /\hbar= & \Delta_{D\text{R}} \sigma_D + (\Delta_{P\text{L}} + \Delta_\text{CL} + \Delta_{D\text{R}})\sigma_{P} + 	\Delta_{SS'}\sigma_{S'} \nonumber \\  
			& + \Delta_\text{CL}\, a^\dagger a \label{eq:H_I}  \\ 
			& + g \left( \sigma_{PD}\, a + \sigma_{DP}\, a^\dagger \right) + \eta (a + a^\dagger) \nonumber \\ 
			&  + \Omega \left(\sigma_{SD} + \sigma_{DS}\right). \nonumber
	\end{align}
	Here, $\Delta_{D\text{R}}=\omega_D - \omega_\text{R}$ is the detuning between the Ramsey spectroscopy laser and the $\ket{D}$--$\ket{S}$ transition, $\Delta_{P\text{L}}=\omega_P - \omega_\text{L}$ is the detuning between the cavity drive laser and the $\ket{P}$--$\ket{D}$ transition, $\Delta_\text{CL}=\omega_\text{C}-\omega_\text{L}$ is the detuning between the cavity drive laser and the cavity mode, and $\Delta_{SS'}$ is the detuning between state $\ket{S}$ and the dark state $\ket{S'}$.
In the cavity drive term, the drive amplitude $\eta$ for coherent driving on resonance, i.e., for $\Delta_\text{CL}=0$, is given by $\eta=\kappa\sqrt{n_\text{coh}}$ with $n_\text{coh}$ the mean photon number and $2\kappa$ the decay rate of the cavity photons. 
This relation can be derived from the Heisenberg-Langevin equation for the cavity field $a$ in steady state (cf. following section).
The value used of $\eta$ in the simulation stems from the calibration of the mean photon number.

% Master equation
\subsection{Master equation}
The system evolution is calculated by numerically integrating the following master equation in Python, using QuTiP~\cite{Johansson12,Johansson13}. 
The master equation consists of four terms, describing unitary evolution, atomic decay, cavity decay with rate $\kappa$, and incoherent cavity driving, derived from a stochastic drive term~\cite{Gardiner94,Gardiner04s}: 
	\begin{align}
		\frac{d\rho}{dt}=&-\frac{i}{\hbar}\left[\rho, H_I\right] \label{master_equation} \\ 
	& +\sum_{i=D,S,S'} \frac{\Gamma_i}{2} \left( 2\sigma^-_{i}\rho \sigma^+_i - \rho \sigma^+_i \sigma^-_i  -  	\sigma^+_i\sigma^-_i\rho \right) \nonumber \\ 
	& + \frac{\kappa}{2}\left(2a\rho a^{\dagger} - \rho a^{\dagger}a  -  a^{\dagger}a\rho \right) \nonumber \\
	&  + \delta n \left( \left[ \left[a,\rho\right], a^{\dagger} \right] +  \left[\left[a^{\dagger},\rho\right], a \right] \right) \nonumber
	\end{align}
The photons are described by a Fock state basis, truncated at $n=9$.
This number is sufficient, since for the measured coherent states the mean photon number is below two, which would correspond to a population of below $2\cdot 10^{-4}$ for the Fock state $\ket{n=9}$.
In the experiment, the incoherent drive is implemented by adding white amplitude noise to the RF-amplitude for the acousto-optic modulator of the cavity drive beam. 
The bandwidth of the frequency generator used for generating the noise reaches from DC to $10$~MHz. 
Since the full cavity linewidth is only $2\kappa=2\pi\times136~\text{kHz}$, this can be considered white noise.

Expanding the last term of Eq.~\ref{master_equation} and combining it with the cavity-decay term, we get:
	\begin{align}
		& \frac{\kappa + \delta n}{2} \left(2a\rho a^\dagger -\rho a^\dagger a - a a^\dagger \rho \right)\nonumber \\ 
		+ 	& \frac{\delta n}{2}\left( 2a^\dagger \rho a - \rho a a^\dagger - a a^\dagger \rho \right),\nonumber
	\end{align}
which corresponds to thermal driving of the cavity~\cite{Carmichael99} with a thermal bath with mean photon number $ n_\text{th} = \delta n/\kappa$.	
Since the coherent and incoherent drive do not interfere, the total mean photon number is given by the sum of the coherent and incoherent contributions as $\langle n\rangle = n_\text{coh} +  n_\text{th}$.

% Second transition
\subsection{Second transition $\ket{D'}$--$\ket{P'}$}
For simulating the second transition, (data in Fig.~2(c) and (e) of the main text), the following parameters need to be changed: $g$ is replaced by $g'$, and due to the different Clebsch-Gordan coefficients, only the following decay channels exist: $\Gamma_{\ket{S'}} = 11/15 \,\Gamma_\text{PD}$ for decay from $\ket{P'}$ to the dark state $\ket{S'}$, and $\Gamma_{\ket{S}} = \Gamma_\text{PS}$ for decay from $\ket{P'}$ to $\ket{S}$. 
$\Gamma_{\ket{D}}$ is replaced with $\Gamma_{\ket{D'}}=4/15\,\Gamma_\text{PD}$ for decay from $\ket{P'}$ to $\ket{D'}$.
Note that for this transition, $\ket{P'}$ has no allowed decay to the second ground state $|4^2$S$_{1/2}, m_{J}=-1/2 \rangle$.
In Eq.~\ref{eq:H_I}, $D$ is replaced by $D'$ and $P$ by $P'$.
Note also that the photon polarization is the same for both transitions $\ket{D}$--$\ket{P}$ and $\ket{D'}$--$\ket{P'}$. 

%%%%% Reconstruction algorithm
\section{Reconstruction algorithm}
In order to reconstruct the photon number distribution in the cavity (Fig.~4 in the main text), we first define a likelihood function~\cite{Lvovsky-RMP09} as
\begin{align}
	 L\left(\eta, \delta n\right) = & \prod_{k=1}^{N}   \left[ P_k\left( \eta, \delta n \right)\right]^{f_k} \\ 
	&  \times  \left[1- P_k\left( \eta, \delta n \right)\right]^{1-f_k} \times \text{const.}  \nonumber
\end{align}
In this formula, $N=51$ is the number of points per Ramsey fringe, $f_k$ is the measured probability to find the ion in $|D\rangle$ for point $k$ in the fringe, $P_k$ is the excitation probability expected from solving the master equation with the cavity drive parameters $\eta$ and $\delta n$ as input, and $\text{const}$ is a scaling factor. 
The quantity $L$ describes the likelihood to observe the measured result (given by $f_k$) for certain parameters $(\eta, \delta\eta)$, based on the model of the system (given by $P_k(\eta,\delta\eta)$). 
The parameters that best describe the data are obtained by maximizing the likelihood or its logarithm
\begin{align}
	\log \left[ L\left(\eta, \delta n\right)\right] = & \sum_{k=1}^{N} \bigl( f_k \log \left[ P_k\left( \eta, \delta n \right)\right]  \label{likelihood} \\ 
	& + \left(1-f_k\right) \log \left[1- P_k\left( \eta, \delta n \right)\right] \bigr) + \text{const.} \nonumber
\end{align}
In order to obtain $P_k$ for a given set $(\eta, \delta n)$, we numerically integrate the master equation Eq.~\ref{master_equation}.
The number $N$ was chosen such that there are a sufficient number of points for the sinusoidal fits to the Ramsey fringes.

The iterative algorithm for maximizing the likelihood function runs as follows:
\begin{enumerate}
	\item Integrate Eq.~\ref{master_equation} for given values of $\eta$ and $\delta n$.
	\item Calculate the likelihood function using Eq.~\ref{likelihood}.
	\item Change $\eta$ and $\delta n$ and integrate Eq.~\ref{master_equation}.
	\item Calculate the likelihood using Eq.~\ref{likelihood} again.
\end{enumerate}
This sequence is iterated until the maximum value of the likelihood has been found in a Nelder-Mead simplex optimization. 
The corresponding values of $\eta_\text{opt}$ and $\delta n_\text{opt}$ are the most likely ones to explain the measured data, and the reconstructed photon number distribution is given by the corresponding diagonal elements $p(n)$ of the cavity density matrix obtained from integrating the master equation with $\eta_\text{opt}$ and $\delta n_\text{opt}$ as input.

%%%%% Photon number calibration
\section{Photon number calibration}
\label{photon_number_calibration}
We independently calibrated the intracavity mean photon number to be able to compare the reconstructed photon number distribution with the expected values. 
Given the probability for a photon to leave the cavity through the output mirror of $p_\text{out}=11(2)\%$, which corresponds to a total photon detection efficiency $\varepsilon=p_\text{out}\times\zeta=4(1)\%$ (including detector efficiency and optical loss in the path efficiency $\zeta$), we calculate an expected count rate of $2\kappa\times p_\text{out}\times\zeta=38(8)$ kHz for the single-photon counting module (SPCM) at the cavity output for a mean photon number of $\langle n\rangle=1$ in the cavity. % see also ELN entry of 20170314.
This rate corresponds to an expected number of counts of $C_0=475(100)$ during the interaction time of $\tau=50~\mu\text{s}$.
We take the cavity field build-up time into account by including a correction factor $c=0.922$, extracted from a simulation, and accordingly get a number of counts of $C_1=C_0/c=515(108)$ for the calibration.
By measuring the output counts $C$, we are thus able to calibrate the mean photon number in the cavity field for a given input power as $\langle n\rangle = C/C_1$.

%%%% Fit model
\section{Fit model}
For analyzing the Ramsey fringes, we use a fit model of the following form:
\begin{align}
	E(\phi) = B + A \cdot \cos\left(\pi \left( \phi - \phi_0 \right) \right) \nonumber
\end{align}
Here, $E(\phi)$ stands for the excitation of the ion to the state $\ket{D}$ (or $\ket{D'}$), $\phi$ for the phase of the second Ramsey pulse with respect to the first one, $A$ is the amplitude of the fringe, $\phi_0$ the fringe shift, and $B$ is the offset of the fringe. 
The contrast is calculated as $A/B$.
We include the spontaneous emission from the excited state $\ket{P}$ by making the offset $B$ dependent on the off-resonant excitation to state $\ket{P}$ (or $\ket{P'}$). 
$B$ therefore has to be recalculated for each value of $\langle n \rangle$ as
\begin{align}
	B = B_{\langle n\rangle=0} \cdot \exp \left( -\Gamma_{\ket{S'}} p_{P^{(')}} \, \langle n\rangle\, \tau \right) \nonumber 
\end{align}
Here, $p_{P^{(')}}=2g^2\langle n\rangle /\left( \Gamma_{\ket{D^{(')}}}^2 + \Delta^2 \right)$ is the probability to off-resonantly excite the ion from $\ket{D}$ ($\ket{D'}$) to state $\ket{P}$ ($\ket{P'}$), and $B_{\langle n\rangle=0}=0.4915$ is the maximum offset achievable for the given coherence time and $\langle n\rangle = 0$.
This number is half of the maximum achievable excitation of a single Doppler-cooled ion. 

The phase offset of $-0.12(1)\pi$ for $\langle n \rangle=0$, obtained from the fit, is due to an AC-Stark shift of the ion levels, caused by the non-zero spatial overlap of the ion wave packet and a laser field at 783~nm used to actively stabilize the cavity length; this field populates a TEM$_{01}$ cavity mode~\cite{Stute12b}. 

%%%% Error analysis
\section{Uncertainty analysis of the reconstructed photon statistics}
The uncertainties of the reconstructed photon number distributions are determined by quantum projection noise~\cite{Itano93} in the Ramsey measurement.
The following method is used to estimate the uncertainties of the reconstructed states shown in Fig.~4 of the main text.
\begin{enumerate}
	\item For a given Ramsey fringe, the maximum likelihood method returns the parameter set ($\eta_\text{opt}$, $\delta n_\text{opt}$), which determines the corresponding photon number distribution.
	\item A Monte-Carlo simulation is executed to obtain a random Ramsey fringe which takes into account the quantum projection noise: 
		The ion populations of this fringe are based on the measured ion populations with additional noise following a binomial distribution with $250$ cycles used in the experiment. 
	\item We reconstruct ($\eta_{i}$, $\delta n_i$) from the Ramsey fringe obtained in Step 2. 
		The index $i$ indicates the iteration number in the Monte Carlo simulation. 
	\item Steps 2 and 3 are repeated until the standard deviations of all calculated numbers ($\eta_i$, $\delta n_i$) have converged according to the criterion that the standard deviation as a function of the number of samples varies less than $5\%$.
		The uncertainties ($\Delta\eta$, $\Delta\delta n$) are then set to the values of the standard deviations.
		The mean of the obtained $\eta_i$ and $\delta n_i$ is $\eta_\text{opt}$ and $\delta n_\text{opt}$.
	\item The upper limit of the reconstructed distribution is given by $(\eta + \Delta\eta$, $\delta n + \Delta\delta n)$, and the lower limit by $(\eta - \Delta\eta$, $\delta n - \Delta\delta n)$. 
		The uncertainties of the mean photon numbers $\langle n\rangle$ and Mandel Q parameters are calculated by propagating these values.
\end{enumerate}

% Fig. S2
\begin{figure}[!t]
	\includegraphics[width=3.0in]{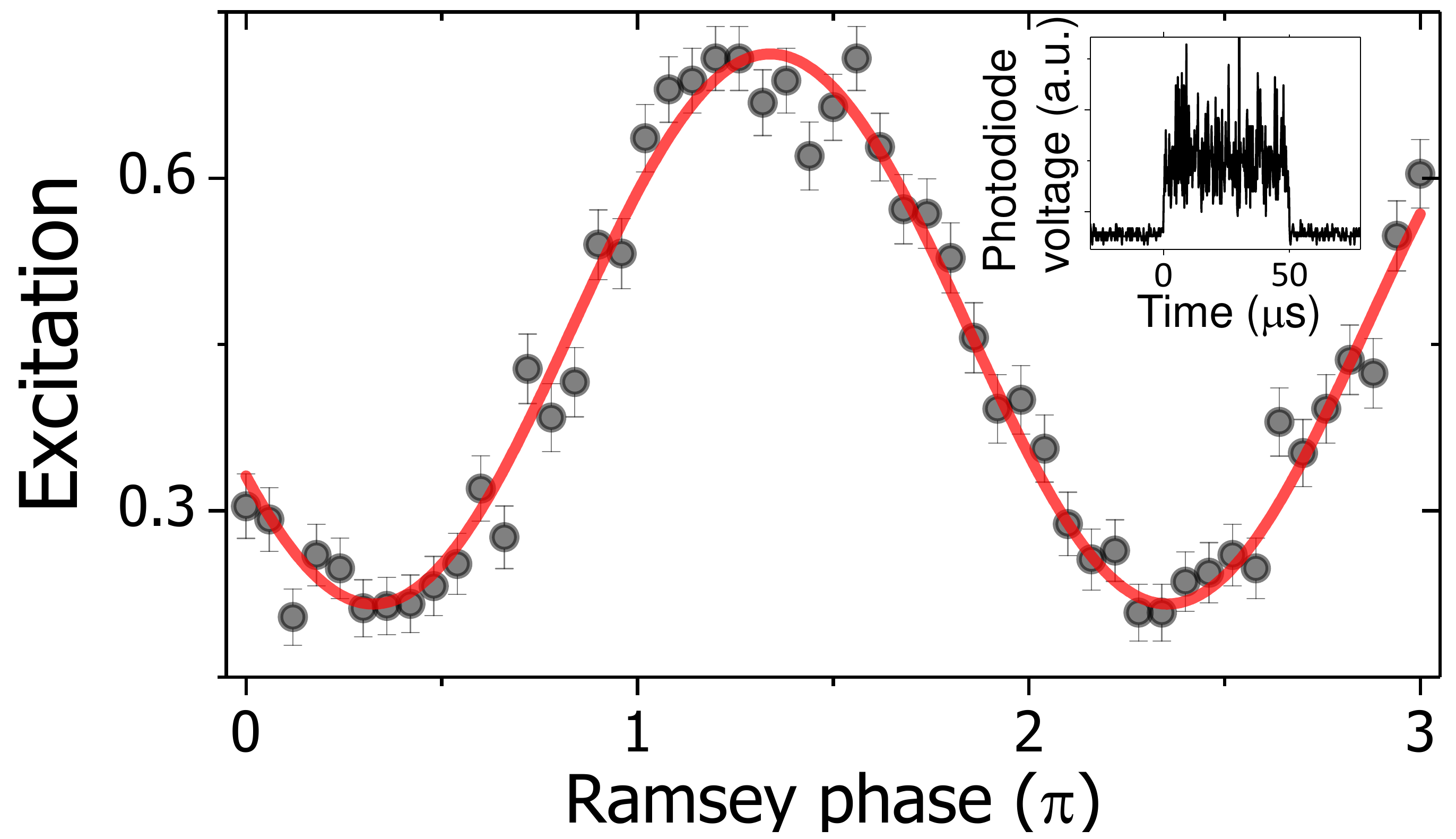} 
	\caption{
		Ramsey fringe when the cavity is driven with a coherent field with additional white noise.
		The black circles are data points and the red line is a sinusoidal fit to the data.
		From the fit, a phase shift of $0.71(2)\pi$ and a contrast of $0.57(3)$ are extracted.
		The error bars are quantum projection noise.
		(inset) Photodiode measurement of the cavity input field.		
	}
	\label{fig:ion_drive_SI}
\end{figure}

%%%% Noise drive
\section{Driving the cavity with additional amplitude noise}
We estimate the intracavity field from a calibration measurement of the cavity drive beam (inset of Fig.~\ref{fig:ion_drive_SI}) with a photodiode.
In the photodiode signal, there is a contribution from the coherent statistics, which is calibrated independently (see Sec.~\ref{photon_number_calibration}), and a noise contribution with thermal statistics.
The coherent statistics is determined by the coherent amplitude $\eta$, while the thermal part is described via the mean thermal photon number $n_\text{th}=\delta n/\kappa$.
By dividing the voltage on the photodiode into the offset part (coherent statistics) and oscillations on top (thermal part), we can calibrate $\delta n$ as a function of the amplitude $V_\text{AC}$ of the oscillations: we first extract the conversion factor $S_\text{V}$ between photodiode voltage and SPCM counts, as $C=S_\text{V} \cdot V_\text{DC}$, where $C$ is the number of measured counts in the SPCM and $V_\text{DC}$ is the voltage measured on the photodiode for coherent driving. 
Next, we can calculate the noise contribution via $\delta n = \kappa S_\text{V} V_\text{AC}/C_1$, using the fact that the number of SPCM counts originating from the thermal part is proportional to $V_\text{AC}$.
In this way, we calculate the mean coherent and thermal photon numbers of the expected as $n_\text{coh}=0.64(14)$ and $n_\text{th}=0.44(9)$, both of which agree with the reconstructed values $n_\text{coh, rec}=0.68(16)$ and $n_\text{th,rec}=0.47(15)$.   

%%%%% Phase resolution
\section{Phase resolution of the Ramsey measurements}
The phase resolution $\delta\phi$ of our Ramsey measurement is limited by quantum projection noise~\cite{Itano93} in the measurement of the ion's state.
In order to estimate $\delta \phi$, we start by simulating a single reference fringe for a mean photon number $\langle n\rangle = 1$ with $N$ phase values, by numerically solving the master equation Eq.~\ref{master_equation}. 
Using the excitation probability $Q_k$ for a given phase $\phi_k$ in this fringe, we draw a random sample $m_k$ with a probability $\Pi_k$ given by the binomial distribution
\[
	\Pi_k(m_k)={ M\choose m_k} Q_k^M \left( 1- Q_k \right)^{M-m_k}, 
\]
where $M=250$ is the number of repetitions for measuring the ion excitation and $m_k$ is the number of times that the ion is found in the excited state $\ket{D}$ out of $M$ trials.
The excitation of the ion to state $\ket{D}$ is then calculated as $p_k=m_k/M$ (corresponding to the values $f_k$ in Eq.~\ref{likelihood}). 
This is repeated for all $N$ phase values $\phi_k$ used in the real measurement.
Next, we extract the phase shift of this simulated fringe, by fitting a sinusoidal function to the simulated data $p_k$ vs. $\phi_k$.
This procedure is repeated $50,000$ times and we obtain a distribution of extracted phase values with a standard deviation $\sigma_\phi$.
We define the phase resolution as $\delta\phi= 2\sigma_\phi$, since this is the minimum distance between two phase distributions that are distinguishable.
A value of $\delta\phi = 0.011\,\pi$ is found. 
Translating this result into a resolution of the mean photon number via 
\[
	\delta\phi = \frac{g^2}{\Delta} \tau \, \delta\bar{n},
\]
we find values of $\delta\bar{n}_{PD}=0.013(5)$ for the $\ket{D}$--$\ket{P}$ transition and $\delta\bar{n}_{P'D'}=0.020(8)$ for the $\ket{D'}$--$\ket{P'}$ transition.
In other words, it is possible to distinguish the phase shift of the qubit Bloch vector for cavity field states, whose mean photon number is different by just $\delta\bar{n}_{PD}$ or $\delta\bar{n}_{P'D'}$, respectively.

%%%%% Strong-pull regime
\section{Strong-pull regime in an optical cavity}
We estimate that it would be possible to reach the strong-pull regime ($g^2/\Delta > \kappa$) in the optical domain with state-of-the-art mirrors.
Ref.~\cite{Rempe92} reports a measurement of high-reflectivity mirrors with transmission $T=5\cdot10^{-7}$ and scattering and absorption loss of $A=1.1\cdot10^{-6}$ per mirror for wavelengths near $850$~nm, corresponding to a finesse of $2\cdot 10^6$.
For a cavity with such mirrors and the length of our cavity of $19.98~\text{mm}$, a photon lifetime of $\tau_\text{c}=42~\mu\text{s}$ is expected, while the one-sided output coupling $T/(2T + 2A) = 16\%$ is comparable to that of our cavity.

The results of this estimation for two atomic species (Table~\ref{tab:strong_pull}) show that $g^2/\Delta$ can be larger than $\kappa$ by a factor of $10.7$ with a single $^{40}$Ca$^{+}$ ion, and $159$ with a single $^{133}$Cs atom on the D2 line. 
For the specific transition $\ket{P}-\ket{D}$ in $^{40}$Ca$^{+}$, the value is $4.3$ due to the Clebsch-Gordan coefficient.
The atom-cavity detuning is set as $10$ times larger than the atomic linewidth as in the experiment presented in this paper. 
Note that the ratio between $g^2/\Delta$ and $\kappa$ can be further enhanced by taking advantage of collective atom-cavity coupling. 

In this regime, it would become possible to generate nonclassical states in the cavity:
Detecting the state of the ion was achieved with current technology in $10.5~\mu\text{s}<\tau_\text{photon}$ with $99\%$ readout fidelity~\cite{Myerson08, Noek13}.
The projection of the ion's state corresponding to this fast detection should make it possible to collapse the cavity field onto nonclassical states within the photon lifetime.

\begin{table}[!t]
	\caption{Estimation for strong-pull regime in atom-cavity systems.}
	\label{tab:strong_pull}
	\begin{tabular}{l*{4}{c}r} 
		Species     	       && $^{40}$Ca$^{+}$     && $^{133}$Cs \\
		\hline
		Transition             && $|3^2$D$_{5/2}\rangle-|4^{2}$P$_{3/2}\rangle$ && $|6^{2}$S$_{1/2}, F=4\rangle$ \\
			                   &&                     &&	$-|6^{2}$P$_{3/2}, F=5\rangle$ \\	
		Wavelength (nm)        && $854$               && $852$ \\
		$g/2\pi$ (MHz)         && $1.53$              && $2.8$ \\
		$\gamma/2\pi$ (MHz)    && $11.5$              && $2.6$ \\
		$\kappa/2\pi$ (kHz)    && $1.9$               && $1.9$ \\
		$g^2/(\Delta\,\kappa)$ && $10.7$              && $159$ \\
	\end{tabular}
\end{table}

\end{document}